\documentclass[a4paper,fleqn]{cas-dc}

\usepackage[authoryear]{natbib}
\usepackage{graphicx,times}
\usepackage{subcaption}
\usepackage{amssymb,amsmath}
\usepackage{aas_macros}
\let\oldAA\AA
\renewcommand{\AA}{\text{\normalfont\oldAA}}
\usepackage{pdflscape}

\usepackage{threeparttable}
\usepackage{supertabular}
\newcommand{\cref}{\ref}

\def\tsc#1{\csdef{#1}{\textsc{\lowercase{#1}}\xspace}}
\tsc{WGM}
\tsc{QE}

\begin{document}
\let\WriteBookmarks\relax
\def\floatpagepagefraction{1}
\def\textpagefraction{.001}

\shorttitle{Cen X-3 observed by Insight-HXMT}    

\shortauthors{Q. Liu et al.}  

\title [mode = title]{Studying the variations of the cyclotron line in Cen X-3 using Insight-HXMT}  

\tnotemark[<tnote number>] 

\author[1]{Qi~Liu}
\author[1]{Wei~Wang}[orcid=0000-0003-3901-8403]
\cormark[1]
\ead{wangwei2017@whu.edu.cn}
\author[1]{Wen~Yang}
\author[1]{Xiao~Chen}
\author[1]{Hanji~Wu}

\affiliation[1]{organization={Department of Astronomy, School of Physics and Technology, Wuhan University},
city={Wuhan},
postcode={430072}, 
country={China}}           


\credit{}
\cortext[1]{Corresponding author}

\begin{abstract}
We investigate the cyclotron resonant scattering features (CRSFs) of the accreting X-ray pulsar Cen X-3 and significantly detect the 29 keV cyclotron line features in the hard X-ray averaged spectroscopy studies based on the recent Insight-HXMT observations in 2022, when Cen X-3 has X-ray luminosity $L_{\rm X} > \sim 5 \times 10^{37}$ erg\ s$^{-1}$ in the bands of 2 -- 60 keV. We do not find a harmonic line in the average spectra based on different continuum models. We showed that the CRSF energies have no correlation with time or luminosity in the average spectra. In addition, by performing a pulse phase-dependent spectral analysis, we revealed the fundamental line with the centroid energy ranging from 25 to 29 keV with a high significance over the spin phases. The evolution of the cyclotron line centroid energies over pulse phase is similar to the shape of pulse profiles, illustrating a positive correlation between the energy of CRSFs and the pulse phase flux.  \\ 

\end{abstract}


\maketitle
\section{Introduction}
\label{sec:introduction}

X-ray binary pulsars (XRBPs) were first discovered with the Uhuru satellite \citep{1971ApJ...167L..67G}, and they typically possess a compact object (a highly magnetized neutron star) that accretes materials from a low-mass star companion via Roche lobe overflow or from dense stellar wind of a massive star. The accreted matter is funneled through the magnetosphere following the magnetic field lines and forms accretion columns near the magnetic poles of the neutron star, where accretion power is released as the X-ray emissions. Since their discovery, XRBPs have been extensively studied for different science interest, like the spectral properties, radiation characteristics and accretion geometry.  

The persistent accreting X-ray pulsar, Cen X-3, first identified with Uhuru \citep{1971ApJ...167L..67G,1972ApJ...172L..79S}, is an eclipsing high-mass X-ray binary (HMXB). It has a spin period of $\sim$4.8 s and an orbital period of $\sim$2.1 days (also see \citealt{2023A&A...675A.135K,2023JHEAp..38...32L}) with a decay rate of $\sim$1.799$ \ \times 10^{-6}$ yr$^{-1}$ \citep{2015AA...577A.130F}. The binary system consists of a neutron star with a mass of $\sim$1.21 $\pm$ 0.21 $M_{\odot}$ and an optical companion O6-8 III supergiant star \citep{1974ApJ...192L.135K,1979ApJ...229.1079H} of mass $\sim$20.5 $\pm$ 0.7 $M_{\odot}$ \citep{1999MNRAS.307..357A} and radius of $\sim$ 12 $R_{\odot}$. The distance to the binary system is estimated to be $\sim$ 8.0 kpc \citep{1974ApJ...192L.135K}. The bright X-ray luminosity ($\sim5 \times 10^{37}$ erg\ s$^{-1}$, \citealt{2008ApJ...675.1487S}) and a secular spin-up trend in this system might suggest the presence of the accretion disk. In addition, evidence for the presence of an accretion disk has been claimed. \cite{1986A&A...154...77T} applied a simple geometric model including an accretion disk to model the optical light curves successfully. And the quasi-periodic oscillations observed at 40 mHz \citep{2008ApJ...685.1109R,1991PASJ...43L..43T,2022MNRAS.516.5579L} also indicate the presence of disk. Other structure like the existence of an accretion wake has been also reported by \cite{2008ApJ...675.1487S}. It is noted that the accretion stream structure for Cen X-3 is complex and still unknown.

The broad-band X-ray spectrum for Cen X-3 is usually characterized by a power law modified by a high energy cut-off \citep{1983ApJ...270..711W,2000ApJ...530..429B,2008ApJ...675.1487S}, an iron emission line, and along with a soft excess below 1 keV. An absorption line around 30 keV out of the continuum spectrum, identified as the cyclotron resonance scattering feature (CRSF), was firstly seen by \cite{1992ApJ...396..147N} with Ginga observations using a Lorentzian high-energy turnover function, but they can not provide firm evidence of that feature due to the sensitivity limit of the detector above the energy $>20$ keV. Later, the presence of the CRSF was confirmed by \cite{1998A&A...340L..55S} with BeppoSAX satellite. The absorption-like features can form in a strong magnetic field, where electrons are quantized in Landau levels, and the electron-photon interaction can generate cyclotron lines by resonant scattering of photons with energies corresponding to the Landau levels. These CRSFs can be used to directly determine the magnetic strength of neutron stars because the energy is related to the magnetic field in the line-forming region. The cyclotron line energy is given by 
\begin{equation}
E_{\mathrm{cyc}} \approx \frac{11.57 \ \mathrm{keV}}{(1+z)} B_{12} ,
\end{equation}
where $B_{12}$ is the magnetic field strength of the neutron star surface, in units of $10^{12}$ G, and $z$ is the gravitational redshift. \cite{1998A&A...340L..55S} estimated the magnetic field of neutron star surface to be (2.4-3) $\times 10^{12}$ G for Cen X-3 by using $E_{\mathrm{cyc}}$ $\sim$ 28 keV .

In addition, \cite{2000ApJ...530..429B} and \cite{2008ApJ...675.1487S} investigated the dependence of CRSF parameters on the pulse phase in Cen X-3 with BeppoSAX and RXTE, respectively. \cite{2021MNRAS.500.3454T} also provided new measurements of the cyclotron line energy using Suzaku and NuSTAR. Since the first detection of a cyclotron line the in Her X-1 \citep{1978ApJ...219L.105T}, the number of XRBPs which show evidence of cyclotron lines has increased to $\sim$ 35 \citep{2019A&A...622A..61S}. However, only one cyclotron line is detected in most X-ray pulsars, so far the sources of higher harmonic lines are still rare. The higher energy CRSF at $\sim 46$ keV (interpreted as the harmonic line) of Cen X-3 has been reported by the early Insight-HXMT observations in 2017 and 2018 \citep{2023MNRAS.tmp...61Y}. Here we study the CRSFs and their variations in Cen X-3 with recent observations in 2022 made by HXMT and compare our results with previous observations. An investigation of a correlation between CRSF energy and time or luminosity is also involved. To investigate the accretion geometry and beaming characteristics in this source, we also perform a pulse phase-resolved analysis. 

In this paper, we report the spectral analysis results for the persistent source Cen X-3 based on the available long-term monitoring observations performed with Insight-HXMT in 2022. The paper is organized as follows. In Section 2, we briefly describe the HXMT observations and data reduction. The analysis of spectral fitting is presented in Section 3, we show a significant cyclotron line at $\sim$29 keV in the phase-averaged spectra but do not find evidence for the harmonic line. In addition, the evolution of CRSFs and spectral parameters over the pulse phases is also involved. In Section 4, we will discuss and summarize the spectral variability of phase-averaged and phase-resolved results. 

\section{Observations and Data reduction} \label{sec:Data}


\begin{table}
    \centering
    \caption{The observation IDs of Insight-HXMT used in this study.}
    \label{tab:ObsIDs}
    \begin{tabular}{l|ccc}
    \hline \hline 
    \multirow{2}{*}{ObsID/ExpID} & \multirow{2}{*}{MJD} & \multirow{2}{*}{Exposure time [s]}  \\ \\
    \hline 
P040425300107 & 59769.518 & 1980 \\
P040425300109 & 59769.788 & 2940 \\
P040425300112 & 59770.233 & 3210 \\
P040425300401 & 59773.652 & 3360 \\
P040425300602 & 59779.559 & 2130 \\
P040425300603 & 59779.692 & 1770 \\
P040425300604 & 59779.852 & 2250 \\
P040425300701 & 59781.079 & 2190 \\
P040425300804 & 59783.888 & 3540 \\
P040425300901 & 59785.258 & 3000 \\
P040425301001 & 59787.175 & 2910 \\
P040425301301 & 59794.198 & 4980 \\
P040425301604 & 59808.990 & 3750 \\
P040425301901 & 59815.498 & 2580 \\
P040425301902 & 59815.647 & 750  \\

    \hline
    \end{tabular} 
\end{table}

The observations of Cen X-3 used in this paper were made by the Hard X-ray Modulation Telescope (Insight-HXMT, \citealt{2020SCPMA..6349502Z}), from July 08, 2022, to August 24, 2022, with a total exposure of $\sim$139 ks. Table \cref{tab:ObsIDs} listed the details of observational information. There are three scientific payloads onboard Insight-HXMT satellite, the High Energy X-ray telescope (HE; 20-250 keV), the Medium Energy X-ray telescope (ME; 5–30 keV), and the Low Energy X-ray telescope (LE; 1–15 keV). The effective areas of 5000 $\rm cm^2$, 952 $\rm cm^2$, and 384 $\rm cm^2$ and a time resolution of 25 $\mathrm{\mu}$s, 276 $\rm \mu s$, and 1 ms, are for HE, ME, and LE, respectively.

To carry out data reduction, the Insight-HXMT Data Analysis Software (HXMTDAS v2.04) is used. We calibrate, screen, and extract high-level products from the events files with the following standard criteria suggested by HXMTDAS user guide (also see \citealt{2021JHEAp..30....1W,2022MNRAS.513.4875C}). Tasks \textsl{he/me/legtigen} are used to generate a good time interval (GTI) file with the pointing offset angle $<0.04^\circ$, the elevation angle $>10^\circ$, the geomagnetic cut-off rigidity $>$ 8 GeV, and the eliminated intervals within the South Atlantic Anomaly passage. Tasks \textsl{he/me/lescreen} are used to screen the data. We then corrected the
arrival times of events to the solar system barycenter using task \textit{hxbary}. Tasks \textit{he/me/lelcgen} are used to extract the light curves with a time resolution of 0.0078125 (1/128) sec.

For the spectral analysis, We used tasks \textsl{he/me/lespecgen} to generate the spectra, and 2–10 keV for the LE, 10–30 keV for the ME, and 30–60 keV for the HE is chosen. The tools \textsl{he/me/lebkgmap} can be used to estimate the background. We used XSPEC 12.11.1 \citep{1996ASPC..101...17A} for the spectral fitting analysis. The uncertainties are estimated by the Markov chain Monte Carlo (MCMC) method with a chain length of 20 000 steps.

\section{Spectral analysis}\label{sec:result}

\subsection{Average spectra and fitting results}
\begin{figure}
    \centering
   \includegraphics[width=.49\textwidth]{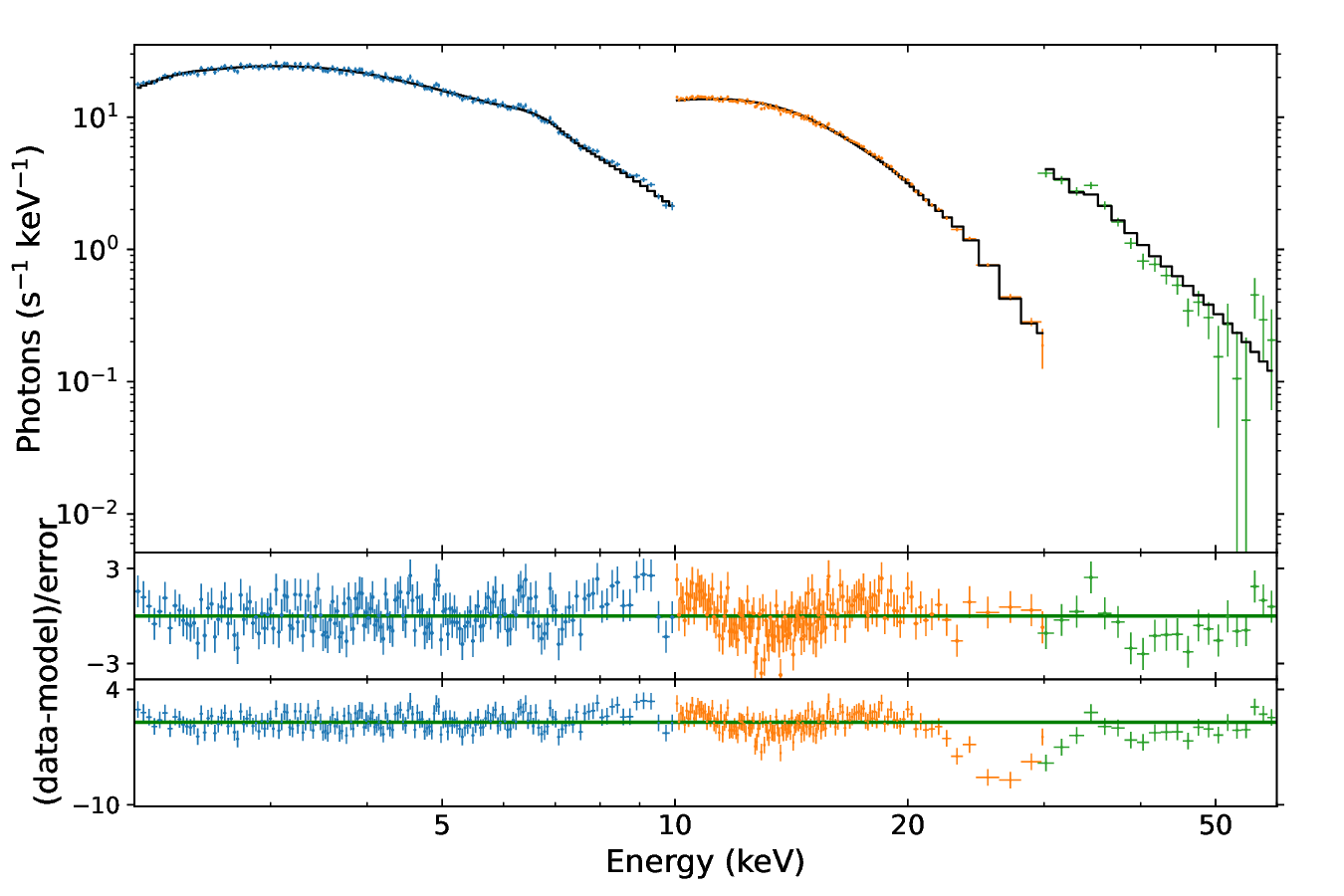}
   \includegraphics[width=.49\textwidth]{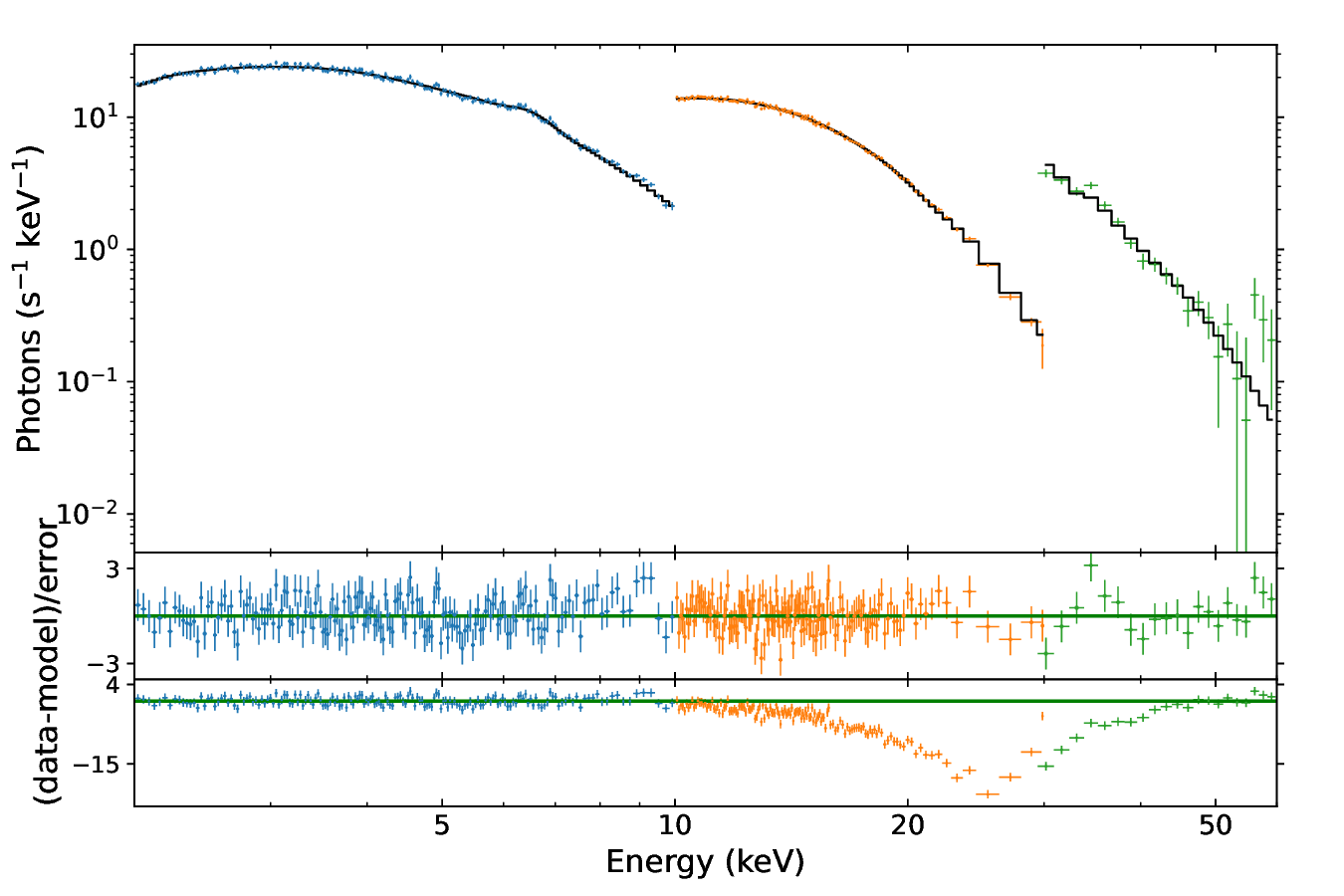}
   \includegraphics[width=.49\textwidth]{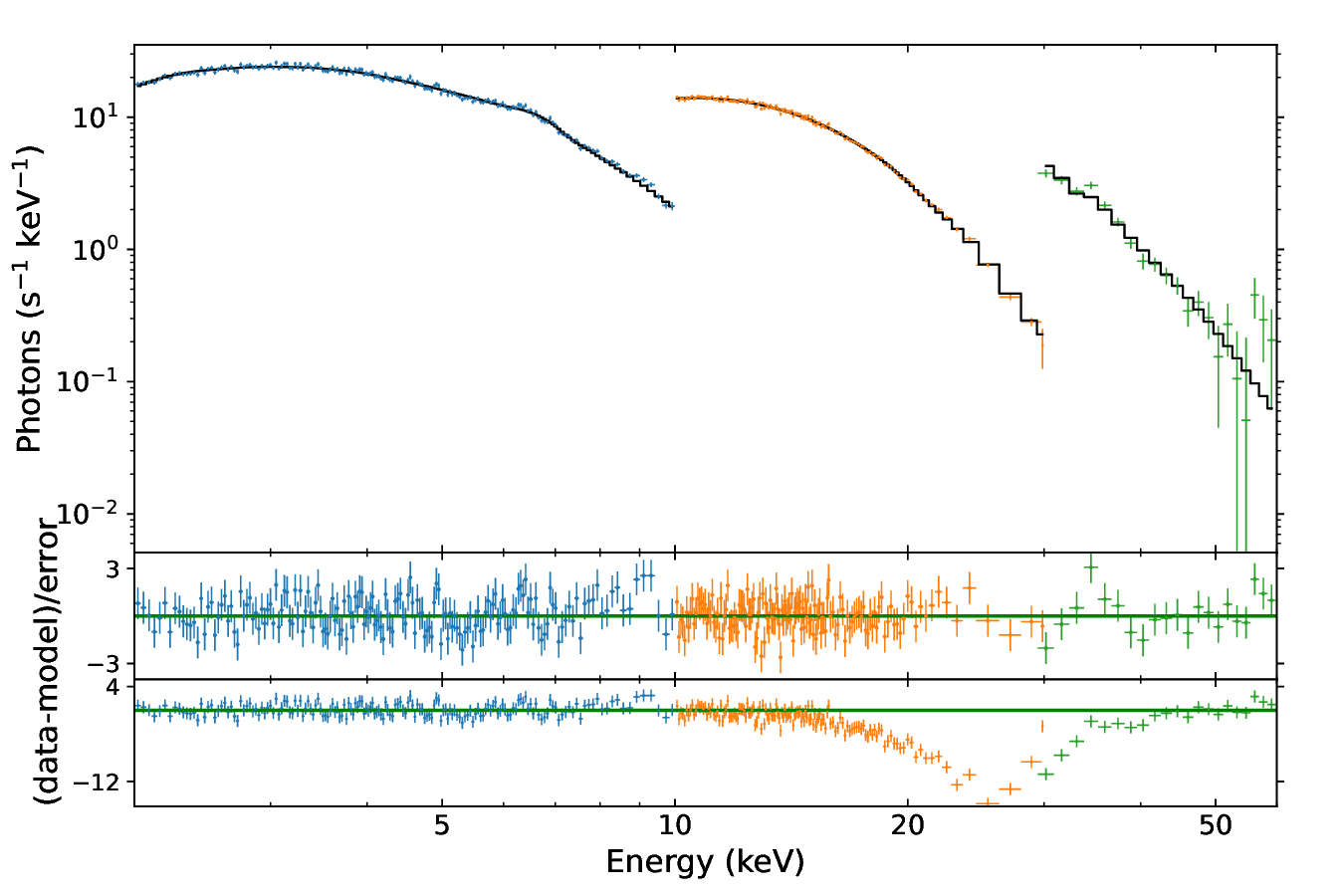}
    \caption{The hard X-ray spectrum of Cen X-3 from 2–60 keV (an example based on ObsID P040425300109) fitted by HIGHECUT (upper panel), NPEX (middle panel
    ), and Fermi-Dirac models (bottom panel). The residuals of the best-fitting model and the best model setting strength of the 30 keV cyclotron line to zero is also plotted.}
    \label{fig:spectrum}
\end{figure}

Here we study the X-ray spectra of Cen X-3 in 2-60 keV. For the continuum spectral analysis, we tried several empirical continuum models which are commonly applied to fit the spectra in accreting pulsars. Here we show the power law with a high energy cut-off model, the Negative Positive Exponential model, and the Fermi–Dirac cut-off power law. A general continuum spectrum of an absorbed power law with a high energy cut-off ($pow * highecut$ in XSPEC model) is usually applied to fit the spectra in Cen X-3 \citep{1983ApJ...270..711W,2000ApJ...530..429B,2022MNRAS.516.5579L}, which is characterized by
\begin{equation}
I(E)=\left\{\begin{array}{ll}
N E^{-\Gamma} & \text { for } E \leq E_{\text {cut }}, \\
N E^{-\Gamma} \times \exp \left[-\left(E-E_{\text {cut }}\right) / E_{\text {fold }}\right] & \text { for } E>E_{\text {cut }} ,
\end{array}\right.
\end{equation}
where $\Gamma$ is the power law index, and $E_{\text {cut }}$ and $E_{\text {fold }}$ are the cutoff and folding energies, respectively. Another form of power law-like continuum, which is a power law with a Fermi-Dirac cutoff, is also employed to fit the
continuum in Cen X-3 \citep{Tamba_2023} and has the following expression
\begin{equation}
I(E)=N \cdot E^{-\Gamma}\left[1+\exp \left(\frac{E-E_{\text {cut }}}{E_{\text {fold }}}\right)\right]^{-1}.
\end{equation}
Finally, the Negative Positive Exponential (NPEX) model, a sum of a positive and a negative power laws multiplied by an exponential cutoff, is also used in the literature \citep{Mihara1995}, which is expressed by
\begin{equation}
I(E)=\left(N_1 \cdot E^{-\Gamma_1}+N_2 \cdot E^{+\Gamma_2}\right) \exp \left(-E / E_{\mathrm{fold}}\right),
\end{equation}
where $\Gamma_2$ is usually set to 2. As it is mentioned, the 30 keV CRSFs have been reported by different missions previously. To model the CRSF, we use a multiplicative absorption line model with a Gaussian optical depth profile (gabs in XSPEC)
\begin{equation}
gabs(E)=\exp \left(-\frac{d_{\mathrm{cyc}}}{\sqrt{2 \pi} \sigma_{\mathrm{cyc}}} e^{-0.5 \left[\left(E-E_{\mathrm{cyc}}\right) / \sigma_{\mathrm{cyc}}\right]^2}\right) ,
\end{equation}
where $E_{\mathrm{cyc}}$ is the cyclotron line energy, with line depth $d_{\mathrm{cyc}}$ and width $\sigma_{\mathrm{cyc}}$. The photoelectric absorption (TBabs in XSPEC) component and iron line emission feature around 6.4 keV (gaussian in XSPEC) are also added in our fitting. In some observations, the width of the iron line is fixed to be 0.2 keV due to poor constraint.

\begin{figure}[!htp]
    \centering
    \includegraphics[width=.5\textwidth]{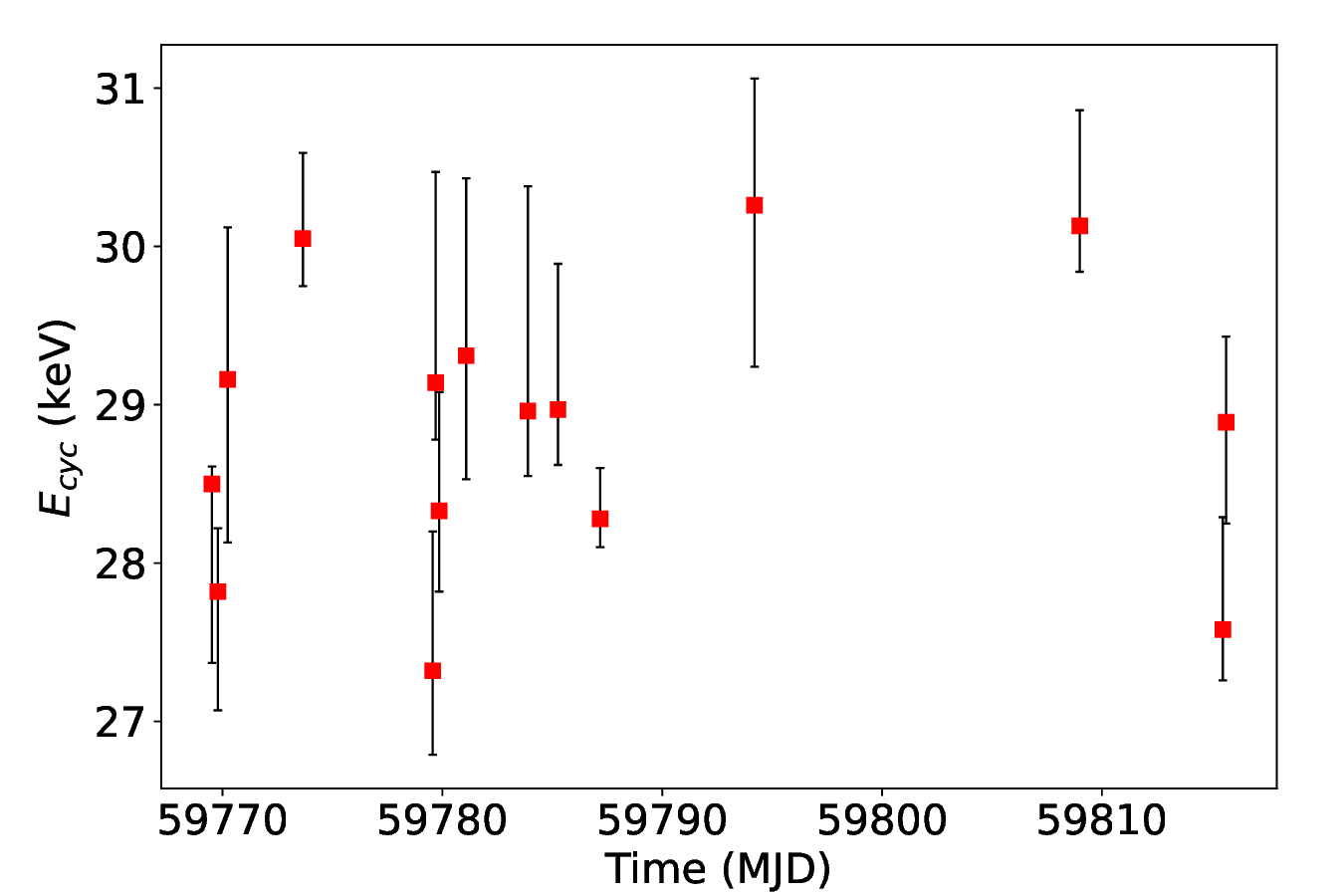}
    \caption{The fundamental CRSF energy $E_{cyc}$ versus the observed time.}
    \label{fig:ratio}
\end{figure}

As an example for the specific observation (e.g., ObsID-P040425300109), we show the fitting results for the three continuum models. The best-fit spectra are shown in Figure \cref{fig:spectrum} and Table \cref{tab:pars} list some best-fit parameters. We found that the cyclotron line parameters do not significantly change for these continuum models. We applied the Fermi–Dirac cutoff power-law continuum model also adopted by \cite{Tamba_2023} in Cen X-3 of NuSTAR data to the spectra for our next analysis. For the results in all available observations in 2022, the centroid energies of the CRSFs for the best fitting are to be found at the mean values of $\sim$ 29 keV and the width of the CRSF varies from 5 keV to 9 keV. All best-fit resulting parameters are listed together in Table \cref{tab:specfit}. We noted that \cite{2000ApJ...530..429B} adopted a blackbody component to describe the soft excess below 1 keV, but this contribution of the blackbody is not markedly detected in our spectra and can be neglected as \cite{2007ApJ...654..435B,2008ApJ...675.1487S} did in their work. In this study, we found that the fundamental CRSF line energy varies from 27 to 30 keV and has no evolution with time (see Figure \cref{fig:ratio}), which is consistent with the previous measurements. And the best-fit residuals in all spectra do not show the significant harmonic line at $\sim$45 keV reported by \cite{2023MNRAS.tmp...61Y} based on the early HXMT observations from 2017 -- 2018. To further check the possible existence of the absorption feature around 40 -50 keV, we tried to add an absorption line with the energy fixed at 45 keV and the width fixed at $\sim$ 10 keV in the fitting of ObsID P040425300604 when the source was most luminous during the observations, and got a line depth of $\sim 1.5 \times 10^{-3}$ without any improvements in $\chi^2$ statistics. Thus, we do not report the obvious 45 keV absorption line in the phase-averaged spectra in the present data.

\begin{table}
\centering
\caption{All the best-fitting parameters in different models for an observation (ObsID-P040425300109) of Cen X-3 observed with Insight-HXMT in 2022. The flux is given in units of erg cm$^{-2}$ s$^{-1}$. For the normalization constants, we fixed the constant to 1 for the LE instrument, and cons\_1 and cons\_2 are the constants referred to for the ME, and HE instruments, respectively.}
\begin{tabular}{lrrr}
\hline
& HIGHECUT & NPEX & Fermi-Dirac  \\

\hline
 $E_{\rm cyc}$ (keV) & $27.83_{-0.26}^{+0.76}$ & $28.50_{-0.71}^{+0.64}$  &   $27.82_{-0.75}^{+0.40}$  \\
 $W_{\rm cyc}$ (keV) & $2.20_{-0.26}^{+0.58}$ & $7.12_{-0.61}^{+0.73}$  &  $6.20_{-0.49}^{+0.34}$   \\
 $D_{\rm cyc}$ & $2.31_{-0.30}^{+0.50}$ &  $10.99_{-1.74}^{+2.34}$  &  $8.47_{-1.22}^{+1.05}$  \\
 $N_{H}$  & $2.70_{-0.11}^{+0.10}$ & $1.66_{-0.16}^{+0.11}$ & $2.09_{-0.12}^{+0.08}$ \\
 $\Gamma$ & $1.39_{-0.02}^{+0.02}$  & $0.55_{-0.04}^{+0.03}$   &  $1.19_{-0.03}^{+0.02}$ \\
 norm\_1 & $0.89_{-0.03}^{+0.03}$ & $0.59_{-0.03}^{+0.03}$  &    $0.74_{-0.02}^{+0.01}$\\
 norm\_2 & ... & $0.0018_{-0.0001}^{+0.0001}$  &   ... \\
$E_{\rm cut}$(keV) & $13.33_{-0.09}^{+0.08}$ &  ... &  $17.85_{-0.82}^{+0.83}$   \\
$E_{\rm f}$(keV) & $10.14_{-0.17}^{+0.17}$ & $4.46_{-0.05}^{+0.06}$  &  $6.86_{-0.17}^{+0.15}$  \\
$E_{\rm Fe}$(keV) & $6.58_{-0.05}^{+0.06}$ & $6.56_{-0.05}^{+0.03}$  &  $6.55_{-0.05}^{+0.03}$ \\
$\sigma_{\rm Fe}$(keV) & $0.37_{-0.05}^{+0.08}$ & $0.29_{-0.05}^{+0.03}$   &  $0.27_{-0.05}^{+0.06}$  \\
$I_{\rm Fe}$(keV) & $0.009_{-0.001}^{+0.002}$ &  $0.007_{-0.001}^{+0.001}$  &  $0.006_{-0.001}^{+0.001}$  \\
cons\_1 & $1.01_{-0.01}^{+0.02}$ & $1.05_{-0.01}^{+0.02}$ & $1.05_{-0.01}^{+0.01}$   \\
cons\_2 & $0.95_{-0.01}^{+0.02}$ & $1.05_{-0.01}^{+0.05}$ & $1.02_{-0.01}^{+0.01}$   \\
reduced $\chi^2$ (dof) & 1.03 (1292) & 0.96 (1292) & 0.96 (1292)  \\
$\rm Log \ Flux_{2-60}$ & $-7.923_{-0.002}^{+0.002}$ & $-7.945_{-0.002}^{+0.001}$ & $-7.939_{-0.002}^{+0.002}$\\
null-hypothesis & 0.236 & 0.852 & 0.857\\
\hline
\end{tabular}
\label{tab:pars}
\end{table}

\begin{table*}[htp]
\centering

\caption{The best-fitting spectral parameters of Cen X-3 in the hard X-ray bands from 2–60 keV. $E_{\rm cyc}$, $W_{\rm cyc}$, and $D_{\rm cyc}$ are the centroid energy, the width, and the depth of for the cyclotron line separately. The flux is given in units of erg cm$^{-2}$ s$^{-1}$, and the absorption column density $N_{\rm H}$ in units of 10$^{22}$ atoms cm$^{-2}$. The other parameters $E_{\rm c}$ (cutoff energy), $\ E_{\rm f}$ (folding energy), $\ E_{\rm Fe}$, and $\sigma_{\rm Fe}$ are in units of keV. Uncertainties are given at the 68\% confidence.}
\label{tab:specfit}

\renewcommand\arraystretch{2.1}
\setlength{\tabcolsep}{0.5mm}{
\begin{tabular}{l|cccccccccccc}
\hline
ObsID & $N_{H}$& $\Gamma$ & $E_{\rm c}$ & $E_{\rm f}$ & $E_{\rm cyc}$ & $W_{\rm cyc}$ & $D_{\rm cyc}$ & $E_{\rm Fe}$& $\sigma_{\rm Fe}$ & Log Flux & $\chi^2$/dof\\ 

\hline 
P040425300107 &  $2.22_{-0.10}^{+0.03}$ & $1.25_{-0.02}^{+0.01}$ & $25.72_{-1.14}^{+1.18}$ & $5.17_{-0.35}^{+0.12}$ & $28.50_{-1.13}^{+0.11}$ & $7.90_{-0.63}^{+0.12}$ & $20.94_{-4.13}^{+2.28}$ & $6.68_{-0.03}^{+0.04}$ & $0.31_{-0.03}^{+0.04}$ & 	$-7.961_{-0.002}^{+0.002}$ & 	1206/1292 \\ \hline 
P040425300109 &	$2.09_{-0.12}^{+0.08}$ & $1.19_{-0.03}^{+0.02}$ & $17.85_{-0.82}^{+0.83}$ & $6.86_{-0.17}^{+0.15}$ & $27.82_{-0.75}^{+0.40}$ & $6.20_{-0.49}^{+0.34}$ & $8.47_{-1.22}^{+1.05}$ & $6.55_{-0.05}^{+0.03}$ & $0.27_{-0.05}^{+0.06}$ & 	$-7.939_{-0.002}^{+0.002}$ & 	1237/1292 \\ \hline
P040425300112 &	$2.13_{-0.11}^{+0.10}$ & $1.15_{-0.02}^{+0.02}$ & $19.20_{-1.01}^{+0.98}$ & $6.38_{-0.32}^{+0.11}$ & $29.16_{-1.03}^{+0.96}$ & $9.10_{-0.63}^{+0.56}$ & $16.67_{-3.24}^{+2.81}$ & $6.59_{-0.02}^{+0.04}$ & $0.20_{-0.05}^{+0.05}$ & 	$-7.942_{-0.002}^{+0.002}$ & 	1156/1292 \\ \hline
P040425300401 &	$2.14_{-0.08}^{+0.09}$ & $1.24_{-0.02}^{+0.01}$ & $27.78_{-0.97}^{+1.20}$ & $5.39_{-0.29}^{+0.25}$ & $30.05_{-0.30}^{+0.54}$ & $8.47_{-0.21}^{+0.34}$ & $28.18_{-2.38}^{+3.37}$ & $6.62_{-0.03}^{+0.03}$ & $0.22_{-0.03}^{+0.05}$ & 	$-7.938_{-0.001}^{+0.001}$ & 	1196/1292 \\ \hline
P040425300602 &	$2.63_{-0.20}^{+0.16}$ & $0.67_{-0.04}^{+0.03}$ & $12.45_{-1.06}^{+1.31}$ & $7.03_{-0.16}^{+0.15}$ & $27.32_{-0.53}^{+0.88}$ & $7.03_{-0.38}^{+0.71}$ & $11.52_{-1.20}^{+2.91}$ & $6.63_{-0.19}^{+0.09}$ & $0.32_{-0.08}^{+0.18}$ & 	$-8.042_{-0.002}^{+0.002}$ & 	1173/1292 \\ \hline
P040425300603 &	$1.80_{-0.15}^{+0.07}$ & $1.05_{-0.03}^{+0.01}$ & $28.40_{-1.46}^{+2.72}$ & $4.65_{-0.52}^{+0.29}$ & $29.14_{-0.36}^{+1.33}$ & $9.17_{-0.19}^{+0.74}$ & $39.28_{-3.30}^{+10.40}$ & $6.78_{-0.06}^{+0.06}$ & $0.13_{-0.07}^{+0.12}$ & 	$-7.907_{-0.003}^{+0.002}$ & 	1175/1292 \\ \hline
P040425300604 &	$1.97_{-0.13}^{+0.18}$ & $1.17_{-0.02}^{+0.02}$ & $27.89_{-1.19}^{+1.84}$ & $4.90_{-0.46}^{+0.25}$ & $28.33_{-0.51}^{+0.75}$ & $8.30_{-0.32}^{+0.38}$ & $29.04_{-3.55}^{+4.98}$ & $6.52_{-0.05}^{+0.08}$ & $0.17_{-0.09}^{+0.07}$ & 	$-7.894_{-0.002}^{+0.002}$ & 	1255/1292 \\ \hline
P040425300701 &	$2.14_{-0.04}^{+0.18}$ & $1.00_{-0.01}^{+0.05}$ & $12.58_{-0.10}^{+1.88}$ & $7.13_{-0.27}^{+0.12}$ & $29.31_{-0.78}^{+1.12}$ & $5.83_{-0.40}^{+0.91}$ & $6.76_{-0.85}^{+2.28}$ & $6.51_{-0.06}^{+0.06}$ & $0.27_{-0.05}^{+0.08}$ & 	$-7.978_{-0.002}^{+0.002}$ & 	1208/1292 \\ \hline
P040425300804 &	$1.54_{-0.08}^{+0.11}$ & $0.90_{-0.02}^{+0.04}$ & $9.78_{-0.27}^{+1.38}$ & $7.31_{-0.17}^{+0.11}$ & $28.96_{-0.41}^{+1.42}$ & $6.72_{-0.35}^{+0.92}$ & $7.00_{-0.73}^{+2.50}$ & $6.54_{-0.03}^{+0.04}$ & $0.19_{-0.04}^{+0.03}$ &  $-7.928_{-0.001}^{+0.001}$ & 	1247/1292 \\ \hline
P040425300901 &	$2.21_{-0.12}^{+0.21}$ & $0.95_{-0.02}^{+0.04}$ & $13.45_{-1.14}^{+1.32}$ & $7.62_{-0.27}^{+0.20}$ & $28.97_{-0.35}^{+0.92}$ & $7.48_{-0.21}^{+0.14}$ & $10.73_{-1.16}^{+1.20}$ & $6.46_{-0.04}^{+0.04}$ & $0.21_{-0.06}^{+0.11}$ & 	$-7.998_{-0.004}^{+0.003}$ & 	1253/1292 \\ \hline
P040425301001 &	$0.00_{-0.00}^{+0.00}$ & $0.76_{-0.01}^{+0.01}$ & $25.23_{-0.92}^{+0.59}$ & $5.10_{-0.12}^{+0.17}$ & $28.28_{-0.18}^{+0.32}$ & $7.45_{-0.13}^{+0.19}$ & $25.27_{-1.80}^{+1.55}$ & $6.47_{-0.10}^{+0.08}$ & $0.20$(fixed) & 	$-8.223_{-0.003}^{+0.003}$ & 	1301/1293 \\ \hline
P040425301301 &	$3.35_{-0.15}^{+0.07}$ & $1.22_{-0.04}^{+0.01}$ & $20.47_{-1.20}^{+1.20}$ & $6.91_{-0.21}^{+0.46}$ & $30.26_{-1.02}^{+0.80}$ & $7.17_{-0.54}^{+0.53}$ & $13.40_{-2.32}^{+3.25}$ & $6.44_{-0.08}^{+0.04}$ & $0.20$(fixed) & 	$-8.011_{-0.002}^{+0.002}$ & 	1434/1293 \\ \hline
P040425301604 &	$1.88_{-0.12}^{+0.05}$ & $1.20_{-0.02}^{+0.01}$ & $30.66_{-1.14}^{+1.41}$ & $4.59_{-0.37}^{+0.29}$ & $30.13_{-0.29}^{+0.73}$ & $7.52_{-0.15}^{+0.36}$ & $27.48_{-2.41}^{+4.44}$ & $6.48_{-0.07}^{+0.06}$ & $0.40_{-0.05}^{+0.08}$ & 	$-8.088_{-0.002}^{+0.002}$ & 	1263/1292 \\ \hline
P040425301901 &	$0.94_{-0.09}^{+0.12}$ & $1.10_{-0.03}^{+0.02}$ & $25.61_{-1.01}^{+0.98}$ & $5.09_{-0.23}^{+0.23}$ & $27.58_{-0.32}^{+0.71}$ & $6.77_{-0.23}^{+0.59}$ & $17.26_{-2.06}^{+3.10}$ & $6.13_{-0.14}^{+0.06}$ & $0.48_{-0.09}^{+0.09}$ & 	$-8.143_{-0.003}^{+0.003}$ & 	1349/1292 \\ \hline
P040425301902 &	$2.09_{-0.19}^{+0.21}$ & $1.23_{-0.03}^{+0.03}$ & $29.20_{-1.75}^{+1.89}$ & $4.75_{-0.52}^{+0.53}$ & $28.89_{-0.64}^{+0.54}$ & $6.61_{-0.27}^{+0.23}$ & $23.99_{-3.96}^{+3.71}$ & $6.37_{-0.03}^{+0.01}$ & $0.21_{-0.06}^{+0.05}$ & 	$-8.120_{-0.004}^{+0.004}$ & 	1330/1292 \\ \hline
\end{tabular} 
}
\end{table*}

\begin{figure}
    \centering
    \includegraphics[width=.5\textwidth]{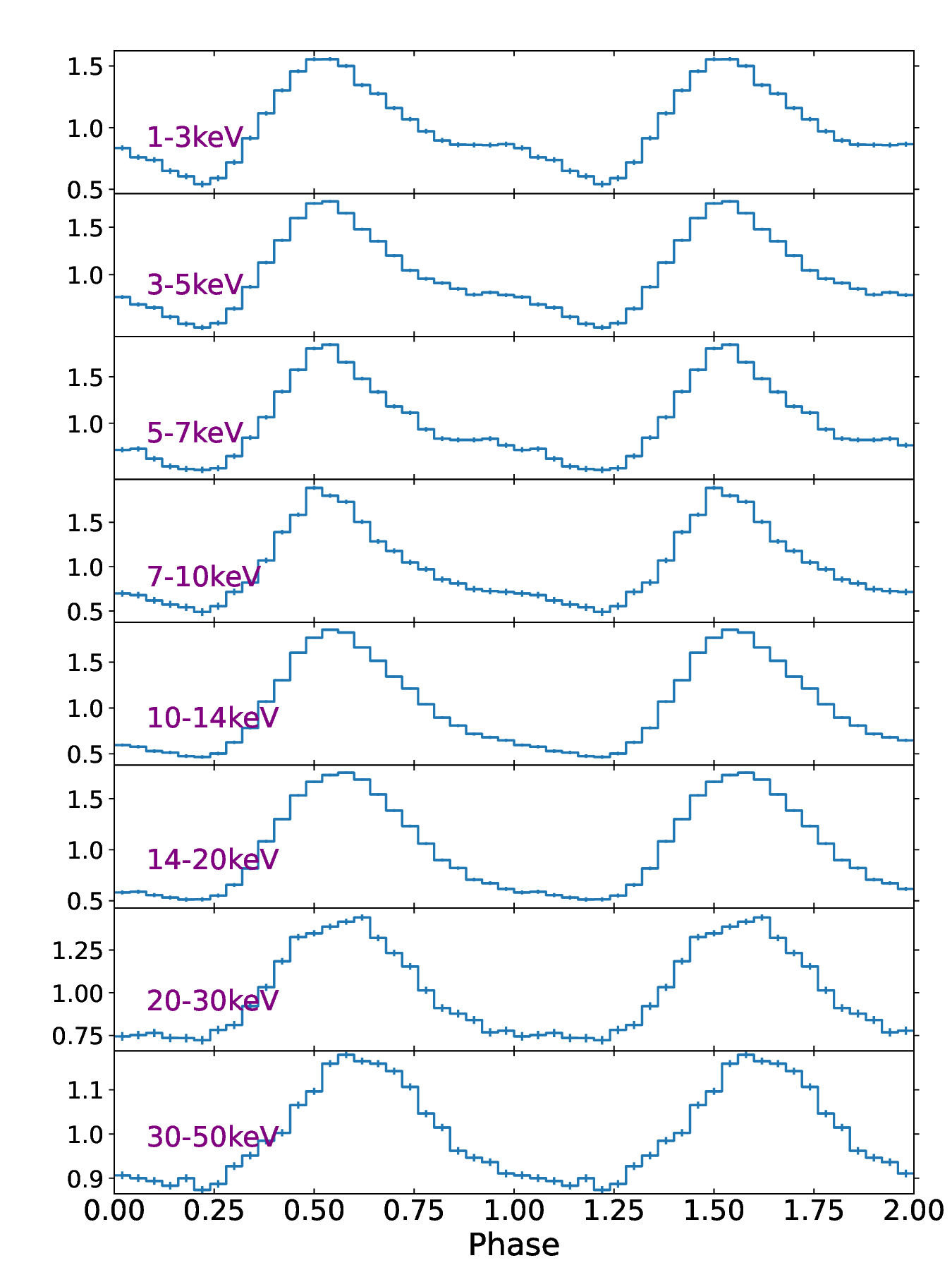}
    \caption{The pulse profiles of Cen X-3 for the ExpID P040425300109 (MJD-59769) in different energy bands based on Insight-HXMT.}
    \label{fig:pulse}
\end{figure}

\begin{figure}
    \centering
    \includegraphics[width=.49\textwidth]{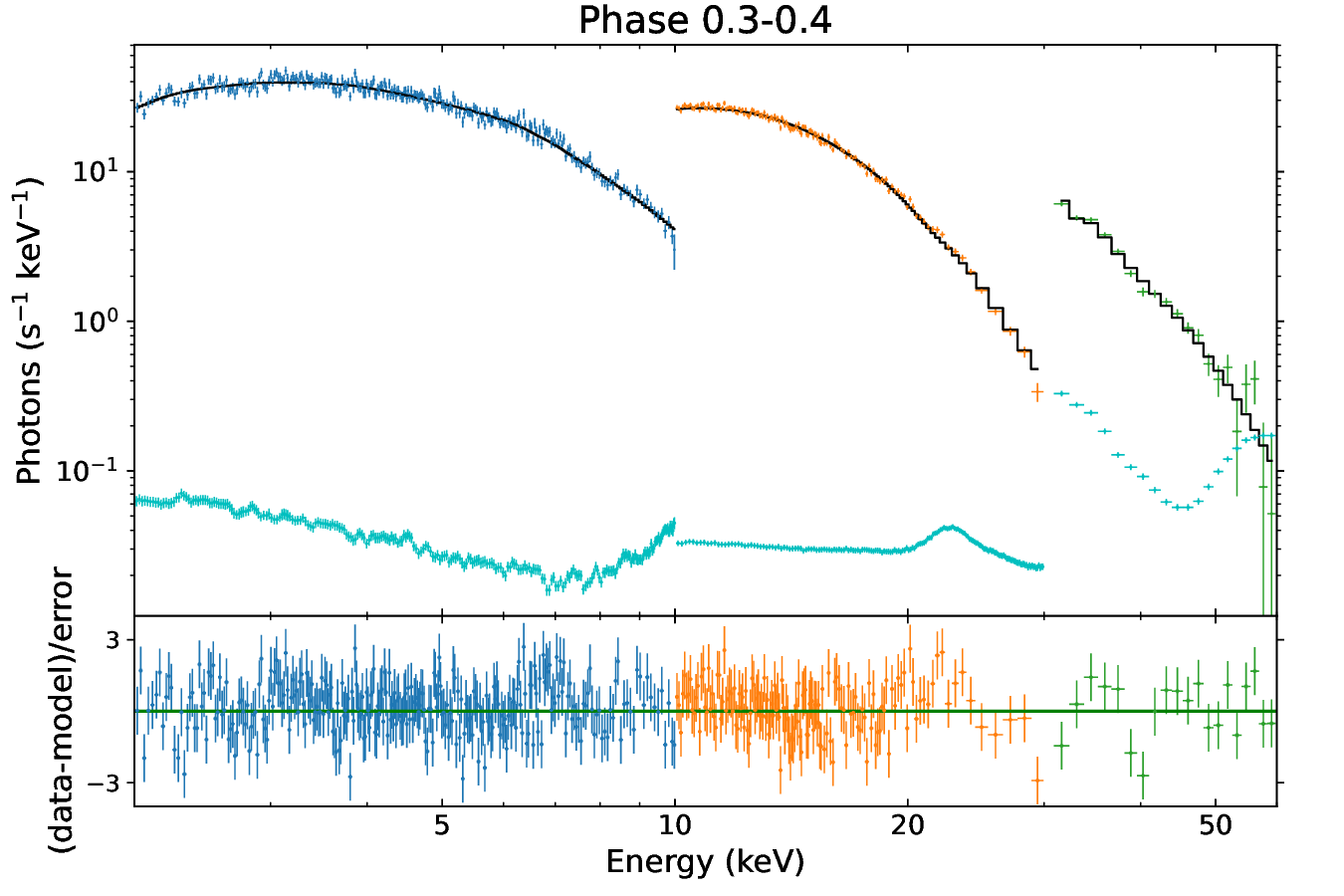}
    \caption{The spectral fitting and residuals of the spectrum combine the observations P040425300109 and P040425300112 at phase 0.3–0.4 in Fermi-Dirac model. The cyan line shows the shape and level of the background flux in the pulse phase, which starts to dominate above $\sim$50-60 keV.}
    \label{fig:phase_spec}
\end{figure}

\begin{figure}
    \centering
    \includegraphics[width=.49\textwidth]{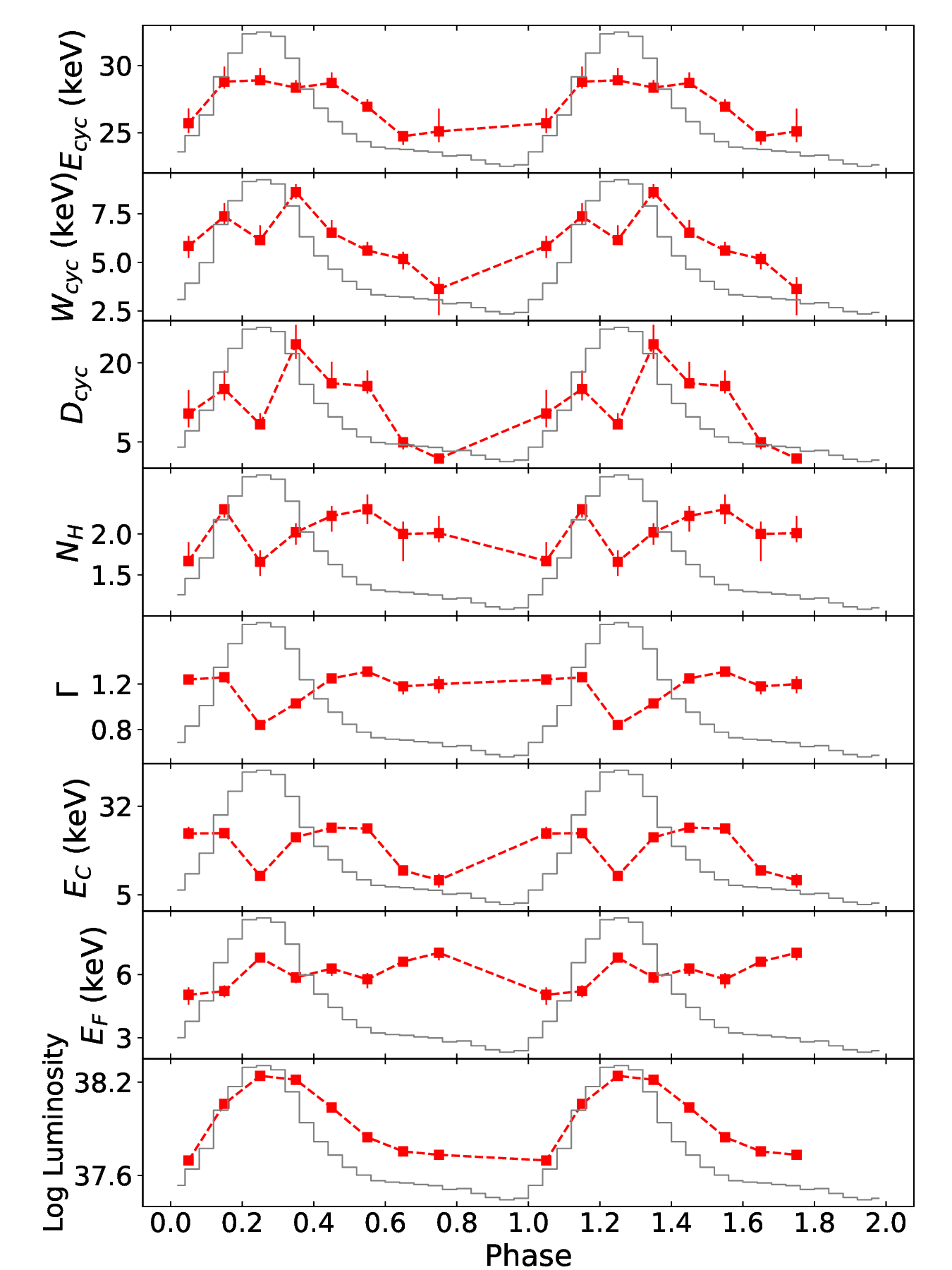}
    \caption{The cyclotron line parameters and the spectral parameters over the phase pulse were shown (the example spectra from the combined spectra of ExpID P040425300109 and P040425300112 observed in one day) in the phase-resolved spectroscopy. The gray line represents the 10-14 keV pulse profile after binary correction.}
    \label{fig:Par_Phase}
\end{figure}

\subsection{Phase-resolved spectra and CRSF detections}

\begin{figure}
    \centering
    \includegraphics[width=.49\textwidth]{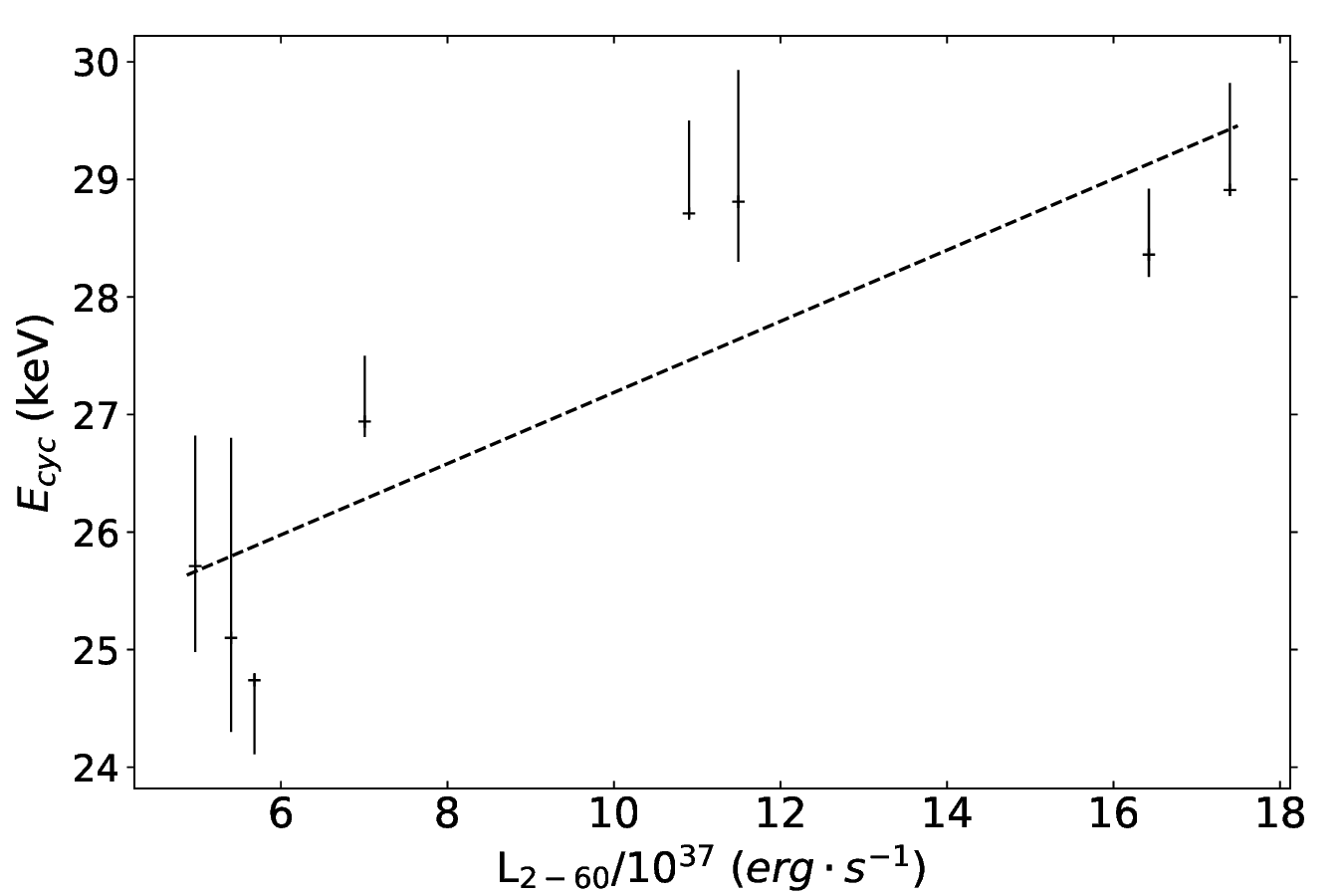}
    \caption{The cyclotron line energies variations as a function of the luminosity were plotted based on the phase-resolved results in Figure \cref{fig:Par_Phase}. The dashed lines are the fitted linear function.}
    \label{fig:E_flux}
\end{figure}

\begin{figure}
    \centering
    \includegraphics[width=.49\textwidth]{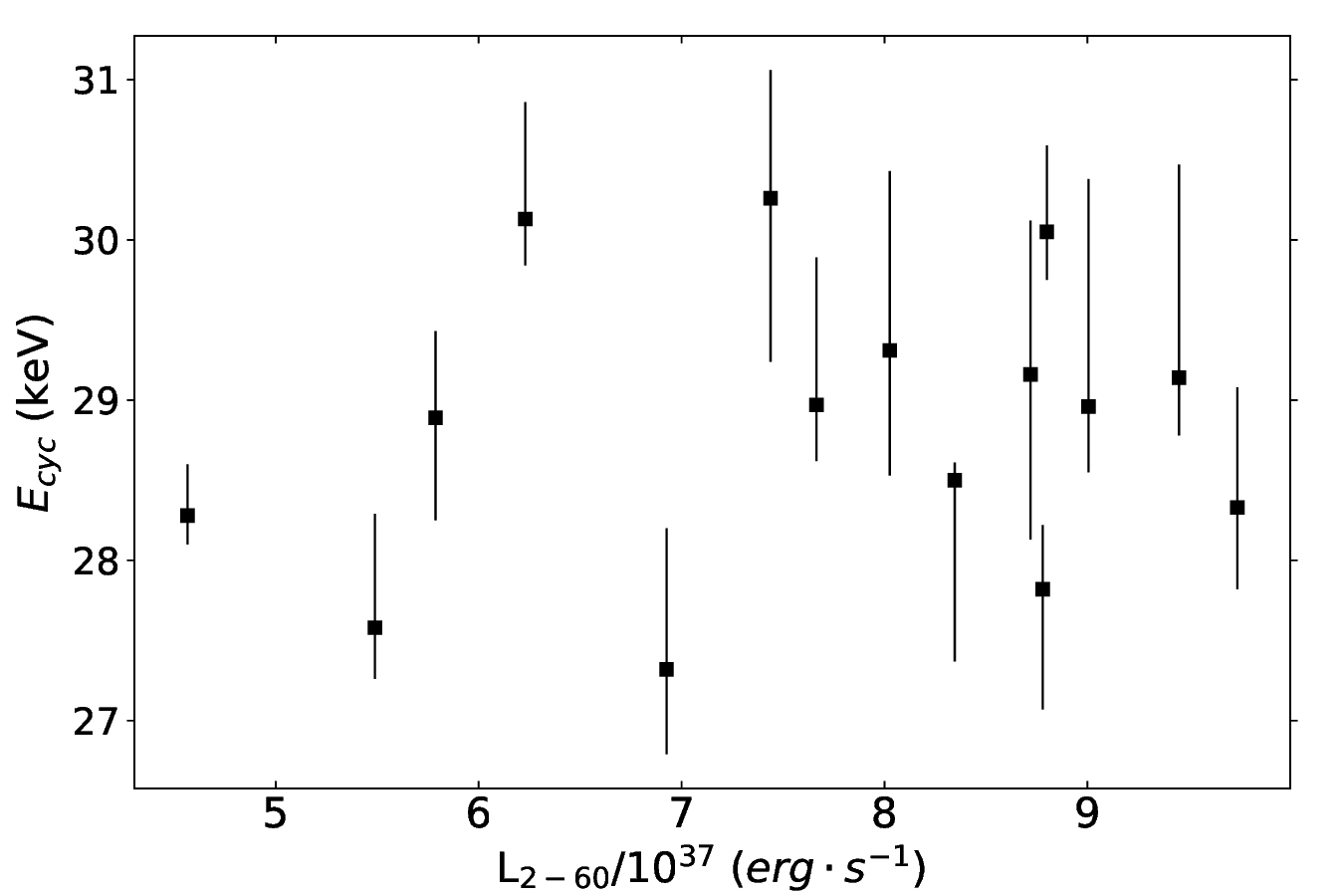}
    \includegraphics[width=.49\textwidth]{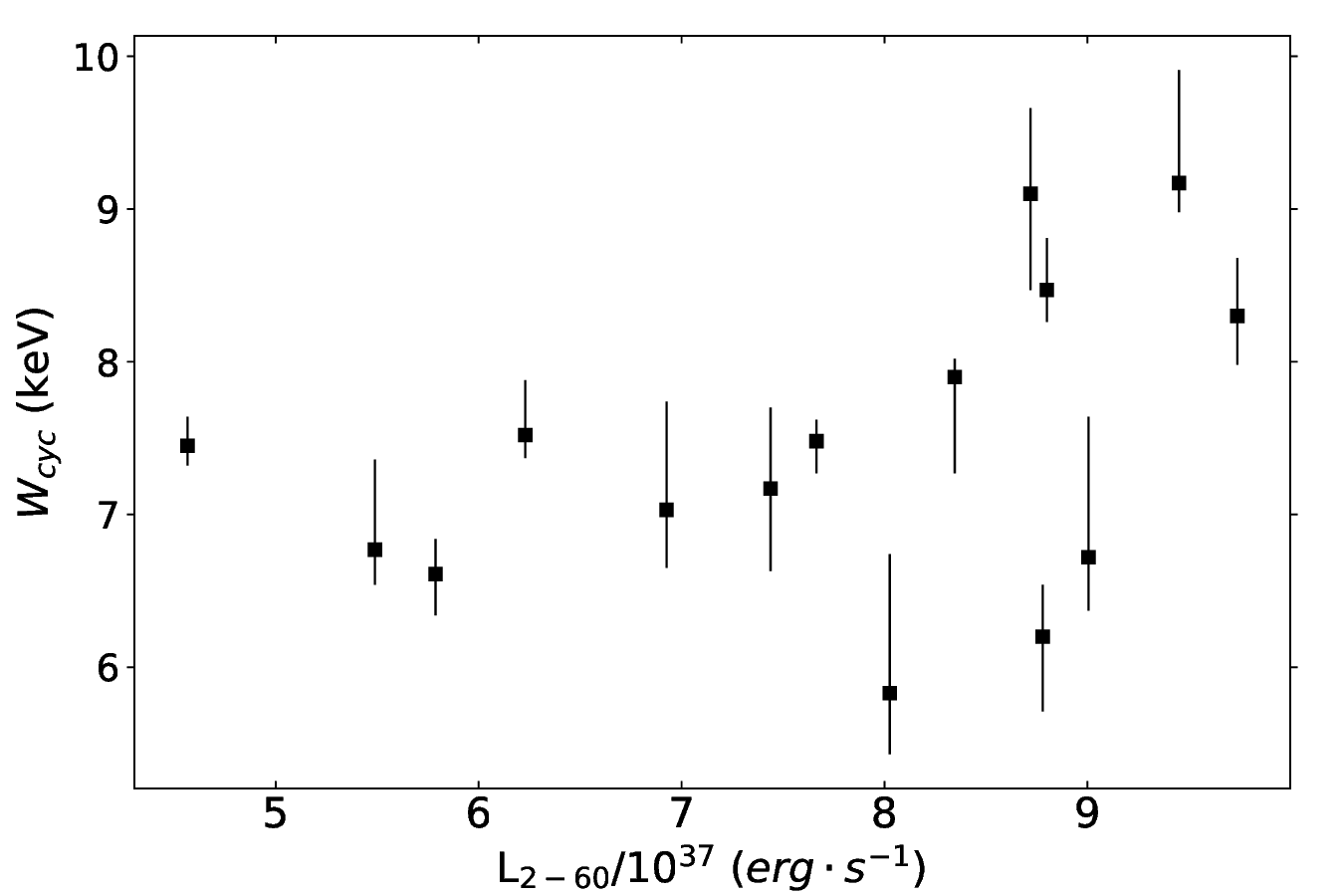}
    \includegraphics[width=.49\textwidth]{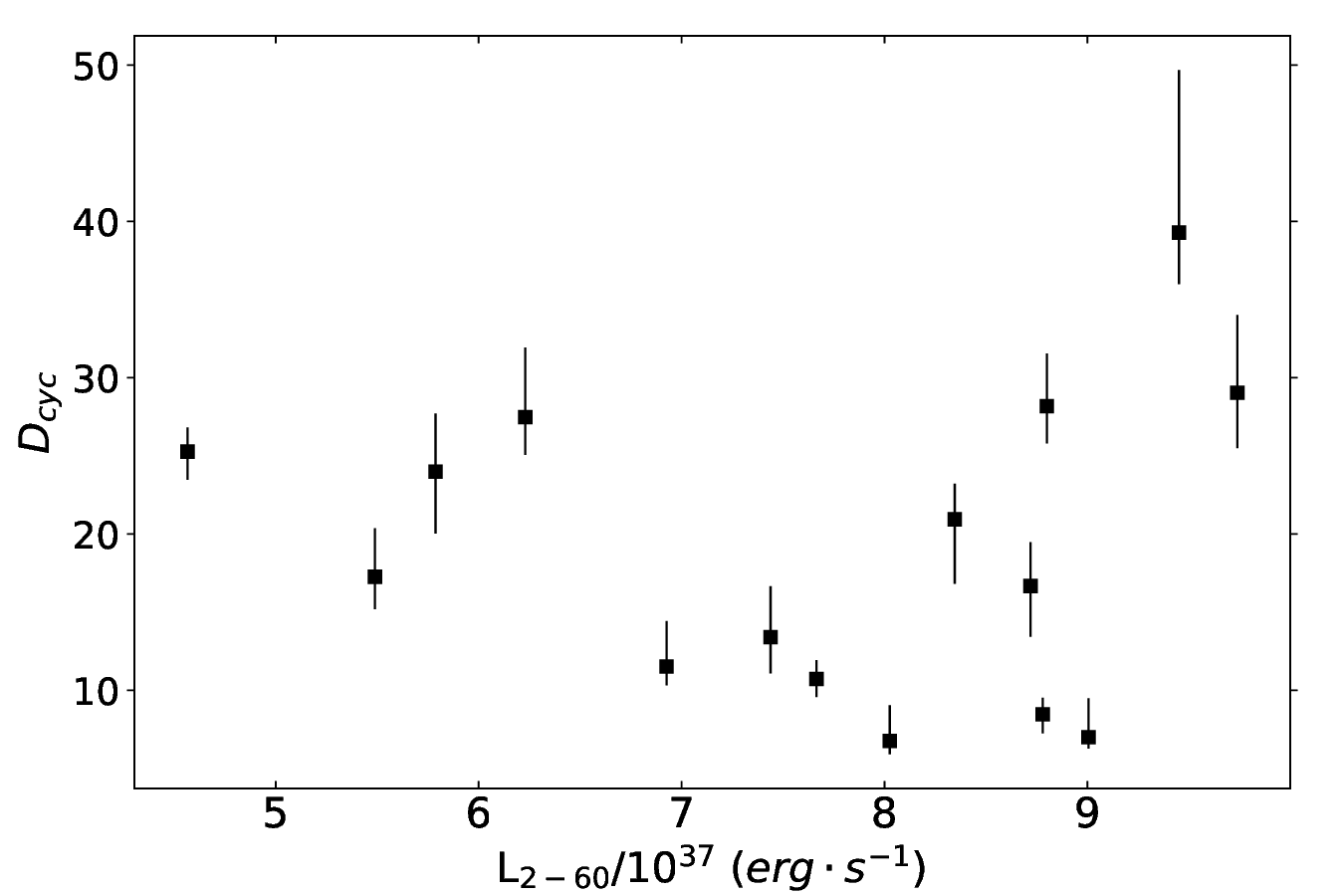}
    \caption{The plots show the CRSFs parameters (energy, width and depth) as a function of X-ray luminosity based on the average spectral analysis in Table \cref{tab:specfit}.}
    \label{fig:GE_Flux}
\end{figure}

For many accreting X-ray pulsars, the centroid energy of the cyclotron line has been found to vary over the different pulse phases (\citealt{2019A&A...622A..61S}). These variations can be associated with the accretion geometry and magnetic field configuration. So we also performed the pulse phase-resolved spectroscopy for Cen X-3. Before generating the phase spectra, we also corrected the pulse arrival times to barycentric times using the orbital ephemeris from Fermi/GBM\footnote{\url{https://gammaray.msfc.nasa.gov/gbm/science/pulsars/lightcurves/cenx3.html}} (i.e., axsin(i):39.653 light-sec; $\rm T_{\pi/2}$:2455073.68504 (JED); orbital period:2.0869953 days; derivative of orbital period:-1.015$\times 10^{-8}$ days/day.). Afterwards, we folded the light curve with the pulse periods derived by \textit{efsearch}. Figure \cref{fig:pulse} shows the energy-dependent pulse profiles over different energy bands generated by folding with the best period as an example (Observation ID P040425300109). We detected a broad pulse with the main peak around the phase of 0.5 in all energy bands, and there was also a very weak or faint second pulse around the phase of 1.0 at a soft energy range of below $\sim$ 10 keV. However, the inter/second-pulse began to fade away at higher energies and the pulse shape became dominated by a single peak. Above $\sim 50$ keV, the pulse feature cannot be clearly observed. We divided the phase into 10 bins to produce the spectra and also generated the new background spectra and response matrices for every phase bin using the HXMT softwares and we ignored 0.8-1.0 phase bins due to the poor statistics in pulse profile minimum (see Figure \cref{fig:Par_Phase}). To improve the data statistics and significance, we combined the consecutive observations of P040425300109 and P040425300112 both of which have a longer exposure of $\sim$ 3000 s using the \textit{addspec} and \textit{addrmf} commands. We noted the phase-resolved spectra in 0-10 phase bins can be fitted by the same model as the phase-averaged well other than the lack of Fe lines because the Fe lines disappear in the continuum probably due to flux statistic fluctuation.   

In the phase-dependent analysis, we also detected the presence of the CRSFs in the different pulse phases of Cen X-3 with a large significance. Figure \cref{fig:phase_spec} shows the spectral fitting and residuals of the spectra at phase 0.3–0.4. We also show the level and shape of the background flux for the pulsed phase. As you can see the cyan line, the background begins to dominate above $\sim$ 60 keV. After successfully performing a fit, with $\chi^{2}$ of 1427 for 1295 d.o.f., the energy of the CRSF line is found to be $28.36_{-0.19}^{+0.56}$ keV, the line depth of $23.51_{-2.75}^{+3.72}$ , which indicate a significant detection of the fundamental cyclotron line around 28 keV ($> 6 \sigma$). Then we presented the variation of the CRSFs and continuum parameters with the pulse phase. As shown in Figure \cref{fig:Par_Phase}, the fundamental line energy reveals the variation from 25 keV to 29 keV along the pulse phases and has higher values in the ascent phases, consistent with \cite{2000ApJ...530..429B,2008ApJ...675.1487S}. The evolution of the fundamental centroid energy over the pulse is similar to the shape of the pulse profile. And the width $W$ of the fundamental CRSFs decreases from 8.5 keV to 3.5 keV around the pulse peak phase. The depth $D$ of the line varies between a value of $2-24$ throughout the phases. 

The $N_H$ remains essentially constant along the pulse profiles, with the averaged value of 2.0 $\times 10^{22} \ cm^{-2}$ and the minimum at the main peak. The photon index $\Gamma$ with an average value of 1.0 shows the variation over the pulse, with a minimum value (harder spectrum) around $\sim 0.2$ at the main peak. The cut-off energy $E_{\rm cut}$ has the similar behavior of $\Gamma$ over the phase, with $E_{\rm cut}\sim 10$ keV near the pulse peak phase, and $E_{\rm cut}\sim 25$ keV around the off-pulse phase. The folding energy $E_{\rm fold}$ does not show a significant variation over the phases with $E_{\rm fold}\sim 5-7$ keV. We also plot the variation of the luminosity over the phases, which is normally consistent with the pulse profile.

As mentioned above, in phase-resolved spectroscopy, both continuum spectral and CRSF parameters show evolution. A significant dependence of the fundamental CRSF energy on the pulse phase is seen. The shape of the changes of the line energies is very similar to the pulse profile. And we also plot the centroid energy of the CRSFs in phase-resolved results as a function of the source luminosity of different phases, as presented in Figure \cref{fig:E_flux}. The centroid energies of the CRSFs show a positive correlation with the luminosity and the Pearson correlation coefficient is about 0.85 with a slope of 0.30 $\pm$ 0.08, and a probability of $\sim$0.007. Here, we define the probability in context, which represents the likelihood under the null hypothesis that there is no correlation in the data. In a lower luminosity ($< \sim 6 \times 10^{37}$ erg s$^{-1}$), they may be negatively correlated with the flux. These changes are generally thought to be caused by the different viewing angles in the line-forming region throughout the rotation of NS. But these changes in the CRSF energy up to 5 keV are not expected. This modulation of the cyclotron line energies over pulse phases also happened among some accreting X-ray pulsars, such as Her X-1 \citep{1982ApJ...263..803V,2013A&A...550A.111V}, Vela X-1 \citep{2022MNRAS.514.2805L}, and GX 301-2 \citep{2004A&A...427..975K}.

\section{Discussion and Summary} \label{sec:discussion}

In this paper, we have reported a detailed spectral analysis for Cen X-3 with observations by HXMT in 2022. We detected the 29 keV cyclotron line features of the source both in the average spectra and the pulse phase-dependent spectra. Next, we will discuss the variability of the spectral parameters in the averaged and resolved spectra, respectively, and then briefly summarize the implications.



From our fitting results (see Table \cref{tab:specfit}) in the continuum in average spectra, we can conclude that the average hydrogen column density $N_H$ is about 2.0 $\times$ 10$^{22}$ atoms cm$^{-2}$, and the photo index varies from 0.8 to 1.3, the cut-off energy is around 13-30 keV. The relatively flat shape at lower energies with photon indices $<2$ can be ascribed to the contribution of thermal Comptonization, and the folding energy at 5-7 keV could be related to the temperature of the Comptonizing plasma in the accretion column \citep{2007ApJ...654..435B}. 

Compared with previous CRSF measurements of the energy range from 29 to 30 keV by BeppoSAX \citep{2000ApJ...530..429B} and RXTE \citep{2008ApJ...675.1487S}, our fundamental CRSF results of 29 keV are consistent with that those of the previous within the uncertainties and have no evident evolution on time. \cite{2019MNRAS.484.3797J} also has shown that the CRSF energy in Cen X-3 is rather stable (approximately 31.6 $\pm$ 0.2 keV) at least the past 14 years with the Swift/BAT data and does not show a long-term decrease like Her X-1 \citep{2015A&A...578A..88K,2016A&A...590A..91S,2017A&A...606L..13S,2020A&A...642A.196S} and Vela X-1 \citep{2016MNRAS.463..185L}.

Many accreting neutron star systems show evidence of the cyclotron lines, and the centroid energy of CRSFs in some X-ray pulsars is found to vary with X-ray luminosity (see a review by \citealt{2019A&A...622A..61S}). The positive correlation of CRSF energy at low or moderate luminosity has been discovered firstly in Her X-1 by \cite{2007A&A...465L..25S}. Other sources like Vela X-1 \citep{2014ApJ...780..133F,2016MNRAS.463..185L,2022MNRAS.514.2805L}, Cep X-4 \citep{2017A&A...601A.126V}, and V0332+53 \citep{2017MNRAS.466.2143D,2018A&A...610A..88V} also show increased CRSF energies with increasing luminosities. And in higher luminosity, a clear negative correlation was observed in V 0332+53 \citep{1990ApJ...365L..59M,2006MNRAS.371...19T}. \cite{2021MNRAS.506.2712D} also found the positive and negative relations for $E_{cyc} - L_x$ in the wind-fed accreting system GX 301-2, showing the critical luminosity around 1.5$\times 10^{37}$ erg s$^{-1}$. During the type II X-ray burst in 2020, 1A 0535+262 \citep{2021ApJ...917L..38K} was confirmed to show an anti-correlation above the critical luminosity and a clear transition between the positive and negative correlation.

At present the critical luminosity has been clearly observed in the transient X-ray binaries V 0332+53 and 1A 0535+262, with more details seen in \cite{2022arXiv220414185M}. Here, we also investigated the cyclotron line energy dependence of the luminosity in Cen X-3. The luminosity is calculated from the fitted flux of the the average spectra (Table \cref{tab:specfit}) based on the source distance of $\sim$8.0 kpc. As shown in Figure \cref{fig:GE_Flux}, we found no significant variations between the CRSF energies and X-ray luminosity for Cen X-3 with the Pearson correlation coefficient of $\sim$ 0.18 and a probability of 0.525. It probably has a decreasing trend at higher luminosities. For further investigating the correlation above 7.0$\times 10^{37}$ erg s$^{-1}$, we fit the relationship of $E_{\rm cyc}- L_x$ with a simple linear function, and obtained Pearson’s correlation coefficient r of $\sim$ -0.43 with a slope of -0.44 ± 0.32 and a probability of 0.209. Therefore, it does not show an evident correlation in higher luminosity ranges. The width of the CRSF shows a possible increasing trend with luminosity, with Pearson correlation coefficients of 0.39 and a probability of 0.153. And the depth of CRSF shows no relation with luminosity, probably decreases below $8\times 10^{37}$ erg s$^{-1}$, and then increases again. Based on the theoretical work \citep{2012A&A...544A.123B}, one can derive the critical luminosity as follows:
\begin{equation}
\begin{aligned}
L_{\text {crit}}\simeq 1.49 \times 10^{37} \ \mathrm{erg} \mathrm{s}^{-1}\left(\frac{\Lambda}{0.1}\right)^{-7 / 5} \times B_{12}^{16 / 15},
\end{aligned}
\end{equation}
for ${M_{\rm NS}}$ =$1.4 M_{\odot}$, ${R_{\rm NS}}$ =10 km, and $\Lambda$ = 1 for the case of spherical accretion, and $\Lambda <$ 1 for disk accretion. Here the surface magnetic field strength $B_{12}$ is computed using the fundamental CRSF energy. For the wind accretion ($\Lambda$ = 1), we obtained a critical luminosity of about 10$^{36} \ \mathrm{erg} \mathrm{s}^{-1}$, which is much lower than the observed luminosities by Insight-HXMT. Then considering the disk accretion case ($\Lambda$ = 0.1), $L_{\text {crit}}\sim 4 \times 10^{37} \ \mathrm{erg} \mathrm{s}^{-1}$, which is slightly lower than the observed X-ray luminosity. But above $L_{\text {crit}}$, a decreasing $E_{\rm cyc}$ with $L_x$ is expected. One possible explanation is that the supercritical luminosity may be even higher ($> 10^{38} \ \mathrm{erg} \mathrm{s}^{-1}$) due to a pretty large polar cap radius or a variety of uncertainties in $\Lambda$.

We briefly discuss the possible scenarios resulting in the phase-dependent evolution of CRSFs. Up to now, the asymmetric variations of cyclotron line energy over pulse phases are still under debate. Regarding the shape of the pulse profile, the asymmetric pulse profile can be explained as the contributions of two emission regions which are displaced from the dipole geometry by approximately 10$^\circ$, as proposed by \cite{1996ApJ...467..794K}. The X-ray polarization research in Cen X-3 performed by IXPE \citep{2022ApJ...941L..14T} showed that the pulsar geometry parameters and the phase-resolved polarimetric properties are well in agreement with the pulse profile decomposition by \cite{1996ApJ...467..794K}. Recently, \cite{Tamba_2023} suggested that a double-peaked shape in low energy and a single-peaked in high energy may indicate the presence of both the pencil and fan beam emissions from the accretion column above the magnetic poles. For the spin-phase variability of CRSF energy, if the variations of cyclotron line energy are due to the height of the accretion column, the change (up to 5 keV) of the line energy ($\sim 20\%$) would require $\sim 1$ km height of the accretion column. These changes could also be produced by a wide polar cap, as suggested by \cite{2000ApJ...530..429B}. \cite{2000ApJ...530..429B} explained the asymmetric variations of the cyclotron line energy by a magnetic dipole offset of 0.1 $R_{NS}$ with respect to the neutron star center. However, \cite{2008ApJ...675.1487S} argued that the assumption by \cite{2000ApJ...530..429B} can only explain the B-field variation seen by BeppoSAX, and is not a viable explanation for higher resolution RXTE data. The magnetic field configuration is possibly rather complicated (such as a multi-pole component for the field). It is therefore not straightforward to interpret the shape of cyclotron line energy. Moreover, the relativistic effects of light-bending are not taken into account. 
The other scenario is the reflection model, as proposed by \cite{2013ApJ...777..115P}. The CRSF features form due to the reflection of X-rays by the atmosphere of the neutron star. The closer the reflected photons are to the magnetic pole, the larger the local magnetic field, resulting in a higher energy of CRSF in the pulse peak.

As a summary, we carried out a detailed spectral analysis from 2 -- 60 keV using the Insight-HXMT observations in 2022, and detected the 29 keV CRSF in the average spectra of the different pointing observations. This CRSF energy shows no significant evolution with the luminosity. And we do not find the existence of a harmonic line in the average spectra. To further study the phase-dependent spectral features, we also found a significant fundamental line around 25-29 keV in different spin phases. And the changes of the CRSF energies over the pulse are similar to the shape of the pulse profile, showing the apparently positive correlation between the CRSF energy and the pulse phase flux. 

\section*{Acknowledgements}
We are grateful to the referee for the useful comments to improve the manuscript. This work is supported by the National Key Research and Development Program of China (No. 2021YFA0718503), the NSFC (12133007). This work has made use of data from the \textit{Insight}-HXMT mission, a project funded by China National Space Administration (CNSA) and the Chinese Academy of Sciences (CAS).

\bibliographystyle{harv}
\bibliography{sample631}{}

\end{document}